\documentclass[5p,twocolumn,11pt]{elsarticle}
\usepackage{amsfonts,color}
\usepackage{amsmath}
\usepackage{amssymb}
\usepackage{graphicx,dsfont}
\journal{Physics Letter B}

\newcommand{\beq}{\begin{equation}}
\newcommand{\eeq}{\end{equation}}
\newcommand{\beqa}{\begin{eqnarray}}
\newcommand{\eeqa}{\end{eqnarray}}

\begin{document}
\begin{frontmatter}
  \title{A digital quantum simulation of the Agassi model}
  
  \author[1,2]{Pedro P\'erez-Fern\'andez}
  \ead{pedropf@us.es}
  
  \author[2,3]{Jos\'e-Miguel Arias}
  \ead{ariasc@us.es}
  
  \author[2,4]{Jos\'e-Enrique Garc\'ia-Ramos\corref{cor1}}
  \ead{enrique.ramos@dfaie.uhu.es}
  
  \author[3]{Lucas Lamata}
  \ead{llamata@us.es}
  
  \address[1]{Dpto.\ de F\'isica Aplicada III, Escuela T\'ecnica Superior de Ingenier\'ia, Universidad de Sevilla, Sevilla, Spain }
  
  \address[2]{Instituto Carlos I de F\'{\i}sica Te\'orica y Computacional, Universidad de Granada, Fuentenueva s/n, 18071 Granada, Spain}

  \address[3]{Departamento de F\'isica At\'omica, Molecular y Nuclear, Facultad de F\'isica, Universidad de Sevilla, Apartado 1065, E-41080 Sevilla, Spain}
  
  \address[4]{Departamento de  Ciencias Integradas y Centro de Estudios Avanzados en F\'isica, Matem\'atica y Computaci\'on, Universidad de Huelva, 21071 Huelva, Spain}

  \cortext[cor1]{Corresponding author at: Departamento de  Ciencias Integradas y Centro de Estudios Avanzados en F\'isica, Matem\'atica y Computaci\'on, Universidad de Huelva, 21071 Huelva, Spain}
  
  \begin{abstract}
    A digital quantum simulation of the Agassi model from nuclear physics
    is proposed and analyzed. The proposal is worked out for the case with four different sites.
    Numerical simulations and analytical estimations are presented to illustrate the feasibility of this proposal with current technology. The proposed approach is fully scalable to a larger number of sites. The use of a quantum correlation function as a probe to explore the quantum phases by quantum simulating the time dynamics, with no need of computing the ground state, is also studied. Evidence is given showing that the amplitude of the time dynamics of a correlation function in this quantum simulation is linked to the different quantum phases of the system. This approach establishes an avenue for the digital quantum simulation of useful models in nuclear physics.
  \end{abstract}
  
\end{frontmatter}

\section{Introduction} 
During the past few decades the possibility of using controllable quantum systems to simulate other quantum systems has been explored extensively~\cite{NoriRMP}. Different quantum platforms have been proposed to reproduce quantum models experimentally, including superconducting circuits, ion traps, cold atoms, quantum dots, as well as quantum photonics~\cite{NoriRMP}. One of the emerging fields proposed for quantum simulations is the analysis of nuclear physics models. In particular, a cloud quantum computing of an atomic nucleus~\cite{LougosvkiNuclear}, quantum simulations of Schwinger-model dynamics~\cite{Zoller1,Zoller2,Zoller3,SavagePRA}, and quantum simulations of quantum field theories with trapped ions and superconducting circuits~\cite{CasanovaPRL,Garcia-Alvarez,MezzacapoPRL,Kihwan} have been proposed and sometimes experimentally realized. For a thorough review of this research field with updated references see Ref.~\cite{BanulsReviewQFT}. However, a paradigmatic quantum nuclear system such as the Agassi model~\cite{Agassi68} has not been analyzed in the context of quantum simulations. Its relevance in Nuclear Physics, but also in a wide variety of fields, including many-body quantum systems and quantum phase transitions, as well as the difficulty to numerically compute the dynamics and static properties of large quantum systems, motivates the quantum simulation of the Agassi model.

The importance of quantum information science (QIS) in Nuclear Physics cannot be underestimated in view of the report of the U.S. Department of Energy (DOE) \cite{Cloe19} where in its Research Opportunity II establishes ``A broad theory program should be supported, which can, e.g., develop methods to address problems in NP using digital quantum computers and quantum simulators, utilize QIS concepts to better understand nuclear phenomena (such as the nuclear many-body problem and hadronization), and develop new QIS applications of importance to nuclear physics''. The present work represents a first step in the direction of this general goal. The use of quantum computing for solving present Nuclear Physics problems has been reviewed in \cite{Zhan21}.

The Agassi model~\cite{Agassi68} is a simple but far from trivial quantum model that includes a combination of long range monopole-monopole and short range pairing interactions. It was first proposed in nuclear physics since it is an exactly solvable model that provides a schematic version of the pairing-plus-quadrupole model that has been extensively used in nuclear structure \cite{PPQ}.  From the quantum phase transition view, this model presents a rich quantum-phase diagram for the ground state, containing several phase transition lines \cite{DaHe86,Garc18,Garc19}, and has been widely studied in a variety of fields. Apart from the symmetric phase, the model has two broken-symmetry phases: one superconducting, linked to the pairing interaction, and another parity-broken phase linked to the monopole-monopole interaction. The phase diagram of the Agassi model has been studied within a mean-field formalism \cite{DaHe86,Garc18,Garc19}. As known, this kind of formalism is valid for the thermodynamic or large-$N$ limit, where $N$ is the number of sites. However, for mesoscopic systems, where finite-$N$ effects are important, the corresponding phases and transitions are blurred and more detailed studies are needed for a clear understanding. In addition, beyond-mean field methods to calculate finite-N effects are difficult to apply
for moderately small-N. For this purpose, quantum platforms could be used to mimic the Agassi model. On the other hand, tools from quantum information, as the quantum discord, have been recently employed to explore the phases in this model to gain insight about its structure \cite{Faba21}. 

In this paper, we propose and analyze the digital quantum simulation of the Agassi model \cite{Agassi68}.
Although we propose a fully digital scheme, for some useful comparisons we refer to trapped-ion platforms \cite{LeibfriedRMP,HaeffnerRPP}. A quantum simulation of the Agassi model may enable one to carry out a full-fledged analysis of this model for a mesoscopic number of sites, in such situations where all classical methods will fail. For instance, apart from the mean field calculations, no finite-N corrections have been calculated for the model in its simplest version,  even for the first correction term. In addition, the extended Agassi model, Ref.\ \cite{Garc18}, includes extra terms producing up to $5$ different phases with three control parameters. 
With our approach, the extension to possible scenarios with inhomogeneous couplings where mean field methods will fail is direct, allowing one for the scalable quantum simulation of nuclear physics models inaccessible to classical supercomputers.
 
\medskip

In this work, we also study how to employ quantum correlation functions as a probe to explore the quantum phases in the system via a quantum simulation of the time dynamics, without needing to compute the ground state. Indeed, we give evidence, analysing the time dynamics of a correlation function, that its amplitude
can be linked to the different quantum phases of the model. Thus, a measure of this time dynamics, that can be done routinely with present technology, will provide the system phase.

\section{The Agassi model}
\label{sec-agassi}
The Agassi model~\cite{Agassi68} consists of $N$ interacting fermions which occupy two levels, each of degeneracy $\Omega$, where $\Omega$ is even, and $j=\Omega/2$. Note that in the following, we will consider $N=2\Omega$. The lower level $\sigma=-1$ has negative parity, and the upper level $\sigma=1$ has positive parity. The magnetic quantum number takes the values $m=\pm 1, \ldots, \pm j$ (note that $m=0$ is excluded). Thus, a single-particle state is labeled by $(\sigma=\pm,m)$. The model is an extension of the Lipkin-Meshkov-Glick \cite{Lipk65} model introduced by D.\ Agassi as a toy model to test many-body theories and to explore the interplay between particle-hole and superfluid correlations. However, the appearance of the model in the literature is scarce. Davis and Heiss \cite{DaHe86} derived the phase diagram of the model and the different collective excitations in the existing phases, using Hartree-Fock-Bogoliuvov (HFB), particle-hole RPA and QRPA approximations. The Agassi model was also used to test some cumbersome numerical methods such as the merging of Coupled Cluster with the symmetry restored HFB theory \cite{Scuseria}. In \cite{Garc18,Garc19}, the authors extended the model by the introduction of new interaction terms that give rise to an extremely rich phase diagram. In \cite{Duke19}, the model was used as a test-bed for a number conserving particle-hole RPA theory. Finally, in \cite{Faba21}, the authors use the Agassi model to study the so called two-orbital quantum discord.
\begin{figure}[tbp]
\centering
\includegraphics[width=0.6\columnwidth]{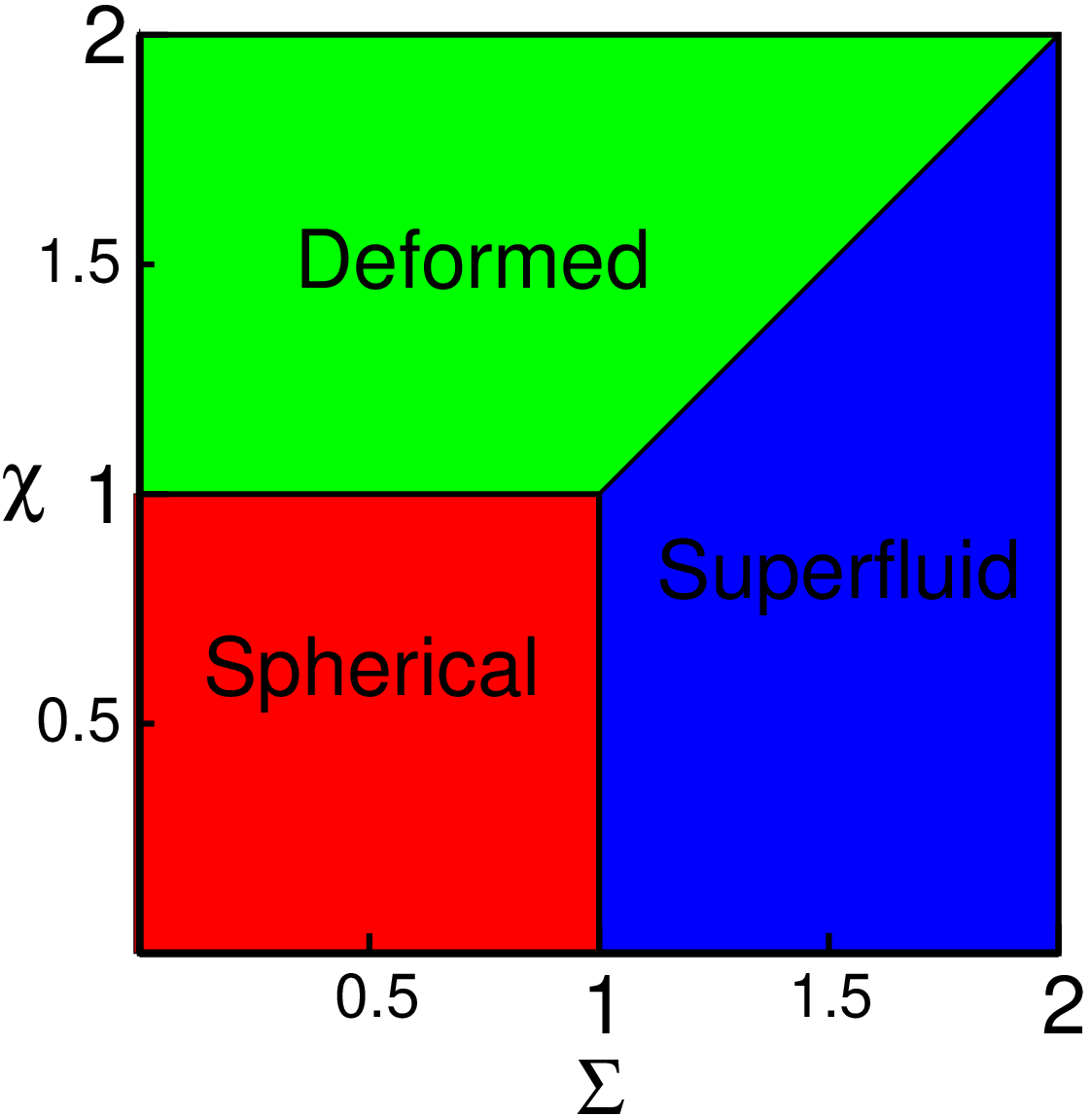}\\
\caption{Phase diagram of the Agassi Hamiltonian (\ref{agassi}), where $\Sigma=\varepsilon g/(2j-1)$ and $\chi=\varepsilon V/(2j-1)$ are scaled control parameters in the Hamiltonian \cite{DaHe86}.}
\label{fig:phasediagram}
\end{figure}

The Agassi Hamiltonian is
\begin{equation}
  \label{agassi}
H=\varepsilon J^{0}-g\sum_{\sigma,\sigma^{\prime}=-1,1}A_{\sigma}^{\dagger} %
A_{\sigma^{\prime}}-\frac{V}{2}\left[  \left(  J^{+}\right)  ^{2}+\left(
J^{-}\right)  ^{2}\right] ~,
\end{equation}
where, implicitly, positive (or null) coefficients are assumed. The operators in $H$ are%
\begin{eqnarray}
J^{+} &=&\sum_{m}c_{1m}^{\dagger}c_{-1m}=\left(  J^{-}\right)  ^{\dagger
} ~,\\
J^{0} &=& \frac{1}{2}\sum_{m}\left(  c_{1m}^{\dagger}c_{1m}%
-c_{-1m}^{\dagger}c_{-1m}\right) ~, \\
A_{1}^{\dagger}  &=& \sum_{m=1}^{j}c_{1m}^{\dagger}c_{1,-m}^{\dagger} = (A_{1})^\dagger ~, \\
A_{-1}^{\dagger} &=&\sum_{m=1}^{j}c_{-1m}^{\dagger}c_{-1,-m}^{\dagger} = (A_{-1})^\dagger ~ ,\\
N_{\sigma} &=&\sum_{m=-j}^{j}c_{\sigma m}^{\dagger}c_{\sigma m},\;\;
N=N_{1}+N_{-1} ~,%
\end{eqnarray}

\noindent where $c^{\dagger}_{\sigma,m}$ ($c_{\sigma,m}$) are fermion creation (annhilation) operators in the state $|\sigma, m\rangle$.

The Agassi model (\ref{agassi}) has a phase diagram with three phases, namely, spherical, deformed, and superfluid, in other words, a symmetric phase and two broken-symmetry phases (deformed HF and superfluid BCS). It is customary to divide the Hamiltonian by $\varepsilon$ and to define the scaled parameters $\chi$ and $\Sigma$ as $g=\frac{\Sigma}{2 j-1}$ and $V=\frac{\chi}{2 j-1}$ to correctly scale the parameters with the system size. The phase diagram of the Agassi model is sketched in Fig.\ \ref{fig:phasediagram}.

\section{Quantum simulation of the Agassi model}
To simulate a quantum model, 
a mapping between the original Hamiltonian and one suited for the
digital simulation
is needed.
Here an Agassi Hamiltonian with $j=1$ is considered. This contains four different sites, to analyze a case that may be experimentally realized with current technology. To simplify the notation we relabel the original fermion operators as
\begin{equation}
c_{1,1}\rightarrow c_1, \ c_{1,-1}\rightarrow c_2, \
c_{-1,1}\rightarrow c_3, \ c_{-1,-1}\rightarrow c_4 ,
\label{eq-As-herm}
\end{equation}
and the corresponding relationships for the creation operators. The mapping is carried out through the Jordan-Wigner image of the above fermions and is written as,
\begin{equation}
c_i=I_1\otimes \ldots\otimes I_{i-1} \otimes \sigma_i^-\otimes \sigma_{i+1}^z \otimes \ldots \otimes\sigma_N^z ,
\end{equation}
and the corresponding Hermitian conjugate one for the creation operator. \(\sigma_i^x, \sigma_i^y, \sigma_i^z\) are Pauli matrices at position $i$ and \(\sigma^\pm_i=\frac{1}{2}(\sigma^x_i \pm {\rm{i}} \sigma^y_i)\). The symbol $\otimes$ stands for the tensor product. We consider in this work $N=4$, then the model space is of dimension $2^4=16$ and, therefore, each operator is given by a $16 \times 16$ matrix. It is worth to mention that the Jordan-Wigner transformation produces  a non-local Hamiltonian.

The spin image of the building block operators is
\begin{eqnarray}
J^+ &=& - \sigma_2^+ \otimes \sigma_3^z \otimes \sigma_4^- - \sigma_1^+ \otimes \sigma_2^z \otimes \sigma_3^- ,\\
J^0 &=& (1/4)( \sigma_1^z + \sigma_2^z - \sigma_3^z - \sigma_4^z) , \label{jzero} \\
  \hspace*{-1.25cm}J^-=(J^+)^\dagger&=& - \sigma_2^- \otimes \sigma_3^z \otimes \sigma_4^+ - \sigma_1^- \otimes \sigma_2^z \otimes \sigma_3^+ ,\qquad\\
A_1^\dagger&=&\sigma_1^+ \otimes \sigma_2^+ , \,\,
A_{-1}^\dagger =\sigma_3^+ \otimes \sigma_4^+ , \\
A_1&=&\sigma_1^- \otimes \sigma_2^- ,\,\,
A_{-1}=\sigma_3^- \otimes \sigma_4^- .
\label{op-mapp}
\end{eqnarray}

Finally, one can to write down the Agassi Hamiltonian (\ref{agassi}) for the case of $j=1$ as,
\begin{equation}
H=H_{1}+H_{2}+H_{3},
\label{H}
\end{equation}
where
\begin{eqnarray}
H_1&=&\frac{\epsilon-g}{4}(\sigma_1^z+\sigma_2^z)-\frac{\epsilon+g}{4}(\sigma_3^z+\sigma_4^z) ,\label{H1}\\
H_2&=& -\frac{g}{4} (\sigma_1^z\otimes \sigma_2^z + \sigma_3^z\otimes \sigma_4^z),
\label{H2}\\
\nonumber H_3&=& -(g+V) (\sigma_1^+\otimes \sigma_2^+ \otimes\sigma_3^-\otimes \sigma_4^-\\
&~& +\sigma_1^-\otimes \sigma_2^- \otimes\sigma_3^+\otimes \sigma_4^+).
\label{H3}
\end{eqnarray}
Note that $H_1$ and $H_2$ only depend on $\sigma^z$ and, therefore, any state with well defined \(\sigma^z\) components will be its eigenstate. $H_3$ depends on $g$ and $V$ and it vanishes for $g=-V$. Moreover, one should consider that,
\begin{equation}
[H_1,H_2]=0,\qquad [H_2,H_3]=0,\qquad [H_1, H_3]\neq 0.
\end{equation}

The term $H_3$ can be further decomposed in terms of tensor products of Pauli matrices, 
\begin{eqnarray}
\nonumber
 \hspace{-1cm} H_3&=&-\frac{g+V}{8}\Big (
         \sigma^{x}_{1} \sigma_{2}^{x} \sigma_{3}^{x} \sigma_{4}^{x}  
         +\sigma^{x}_{1} \sigma_{2}^{y} \sigma_{3}^{x} \sigma_{4}^{y}\\
  \nonumber &+&\sigma^{x}_{1} \sigma_{2}^{y} \sigma_{3}^{y} \sigma_{4}^{x}
                +\sigma^{y}_{1} \sigma_{2}^{x} \sigma_{3}^{x} \sigma_{4}^{y}
                +\sigma^{y}_{1} \sigma_{2}^{x} \sigma_{3}^{y} \sigma_{4}^{x} \\
     &+& \sigma^{y}_{1} \sigma_{2}^{y} \sigma_{3}^{y} \sigma_{4}^{y}
     -\sigma^{y}_{1} \sigma_{2}^{y} \sigma_{3}^{x} \sigma_{4}^{x}   
         -\sigma^{x}_{1} \sigma_{2}^{x} \sigma_{3}^{y} \sigma_{4}^{y} 
         \Big ),\qquad
         \label{H3desarrollado}
\end{eqnarray}
where the symbols $\otimes$ have been taken out to simplify the notation.

It is worth noting that for this simple case, $j=1$, the ion-mapped Hamiltonian (\ref{H}) depends on just one effective control parameter, $g+V$ (see Eq.~(\ref{H3desarrollado})), and not on $g$ and $V$ separately, as in the thermodynamic limit of the model \cite{DaHe86,Garc18,Garc19}. Therefore, it is only possible to distinguish for $j=1$ between a symmetric phase (SP) that is obtained for $g+V<1$ and a broken-symmetry phase (BSP) emerging for $g+V>1$. In this simplest case, the phase diagram is of dimension $1$ as shown in Fig.\ \ref{fig:max_correlation} (upper coloured panel). 
The critical point in the transitional path between these two phases is $g+V=1$. 


\subsection{Theoretical model for the implementation}
In order to carry out a quantum simulation with the Agassi model, we propose to employ a digital protocol, via a Lie-Trotter-Suzuki decomposition~\cite{NoriRMP}. The protocol will rely on expressing the quantum evolution operator $U(t)=\exp(-iHt)$ for the Hamiltonian $H$ in Eq.~(\ref{H}) by means of a Trotterized dynamics, in terms of $H_{1,2,3}$ of Eqs.~(\ref{H1}), (\ref{H2}), and (\ref{H3}),
\begin{equation}
    U(t)\simeq\{ \exp[-i(H_1+H_2)(t/n_T)]\exp[-iH_3 (t/n_T)]\}^{n_T},
\end{equation}
where the error produced will depend on the commutator $[(H_1+H_2),H_3]$ and scale as $1/n_T$, where $n_T$ denotes the number of Trotter steps.

Once the dynamics has been decomposed into the previous blocks, each of these can be implemented efficiently with trapped-ion systems. The operator $\exp(-iH_1t)$ consists of single-qubit gates which are customary with trapped ions, to fidelities often above $99.99\%$ \cite{DavidLucasSingleQubit}. The operator $\exp(-iH_2t)$ is composed of two two-qubit gates which can be carried out via M{\o}lmer-S{\o}rensen gates with fidelities above 99.9\% in some experimental setups~\cite{DavidLucasTwoQubit}, in addition to single-qubit gates to rotate the basis from $x$ to $z$. And finally, $\exp(-iH_3 t)$ consists of the exponential of sum of tensor products of four Pauli matrices, which can be carried out efficiently with trapped ions~\cite{Muller11, Casanova12}. Namely, each exponential of a tensor product of four Pauli operators can be implemented via two M{\o}lmer-S{\o}rensen gates and a local gate, together with the necessary single qubit gates to rotate the bases of the Pauli operators in the tensor product to those needed. Given that all the 4-body terms in Eq.~(\ref{H3desarrollado}) commute, they may be carried out sequentially without digital error, namely, with one Trotter step for the whole $H_3$ term. 

The scaling in our protocol is efficient, given that the number of necessary elementary gates in trapped ions, i.e., single and two-qubit gates, is polynomial in the number of interacting fermions, $N$, in the Agassi model. On the other hand, with a classical computer the scaling would be inefficient given that the Hilbert space dimension would grow exponentially in $N$. Of course, under highly symmetric configurations one may obtain a solution, in some cases, in terms of polynomial resources. However, in general terms, with a generalized Agassi model with, e.g., inhomogeneous couplings, this will not be possible and a quantum simulator such as the one proposed here  will provide an exponential gain in resources with respect to a classical computer.

\begin{figure}[hbt]
  \centering
  \includegraphics[width=1\linewidth]{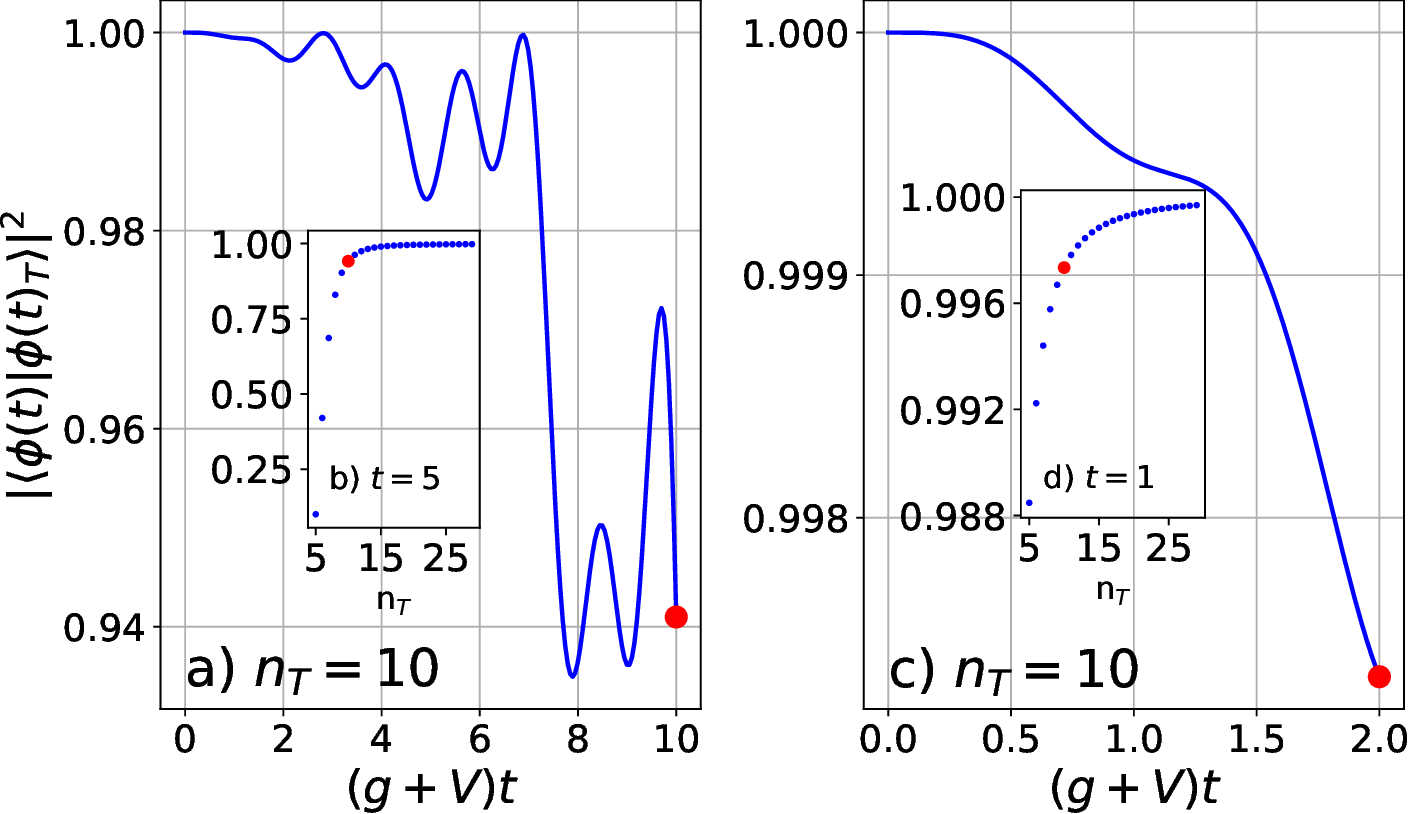}
  \caption{Fidelity $|\langle \phi (t)| \phi(t)_T \rangle|^2$ as a function of $(g+V) t$ for $n_T=10$ in panels a) and c) and as a function of $n_T$ for $t=5$ in panel b) and $t=1$ in panel d). In all cases the parameters of the Hamiltonian are $\epsilon=1$ and $g=V=1$. Red dots in panels a), and b) and also those in c) and d) correspond to the same data points.}
    \label{fig:fidelity}
  \end{figure}

\subsection{Numerical simulations}
Note that for all the calculations presented in this section a certain initial state is considered, in our case, $|\downarrow_1\otimes\downarrow_2\otimes\uparrow_3\otimes\uparrow_4\rangle$ which corresponds to the state with the minimum value of the angular momentum projection, $J^0=-1$ (see definition of $J^0$ (\ref{jzero})).

\medskip

We plot in Fig.~\ref{fig:fidelity} our numerical results for the digital decomposition. In Figs.~\ref{fig:fidelity}a we show the fidelity $|\langle \phi (t)| \phi(t)_T \rangle|^2$ as a function of $(g+V) t$ with $n_T=10$, where $|\phi(t)\rangle$ and $|\phi(t)_T\rangle$ denote the exact numerical state and the one obtained via the Trotterized digital dynamics, respectively. Panel c)  is just a zoom of a) for small $t$.  In Figs.~\ref{fig:fidelity}b and \ref{fig:fidelity}d we depict the fidelity $|\langle \phi (t)| \phi(t)_T \rangle|^2$ as a function of $n_T$ for $t=5$ and $t=1$, respectively, where $t$ denotes the total simulated time interval. The red dots in panels b) and d) correspond to the red dots in panels a) and c), respectively. This figure makes clear that the Trotter dynamics match very efficiently the exact calculation in a large time interval even for small number of Trotter steps, $n_T$. 
\begin{figure}[hbt]
  \centering
  \includegraphics[width=1\linewidth]{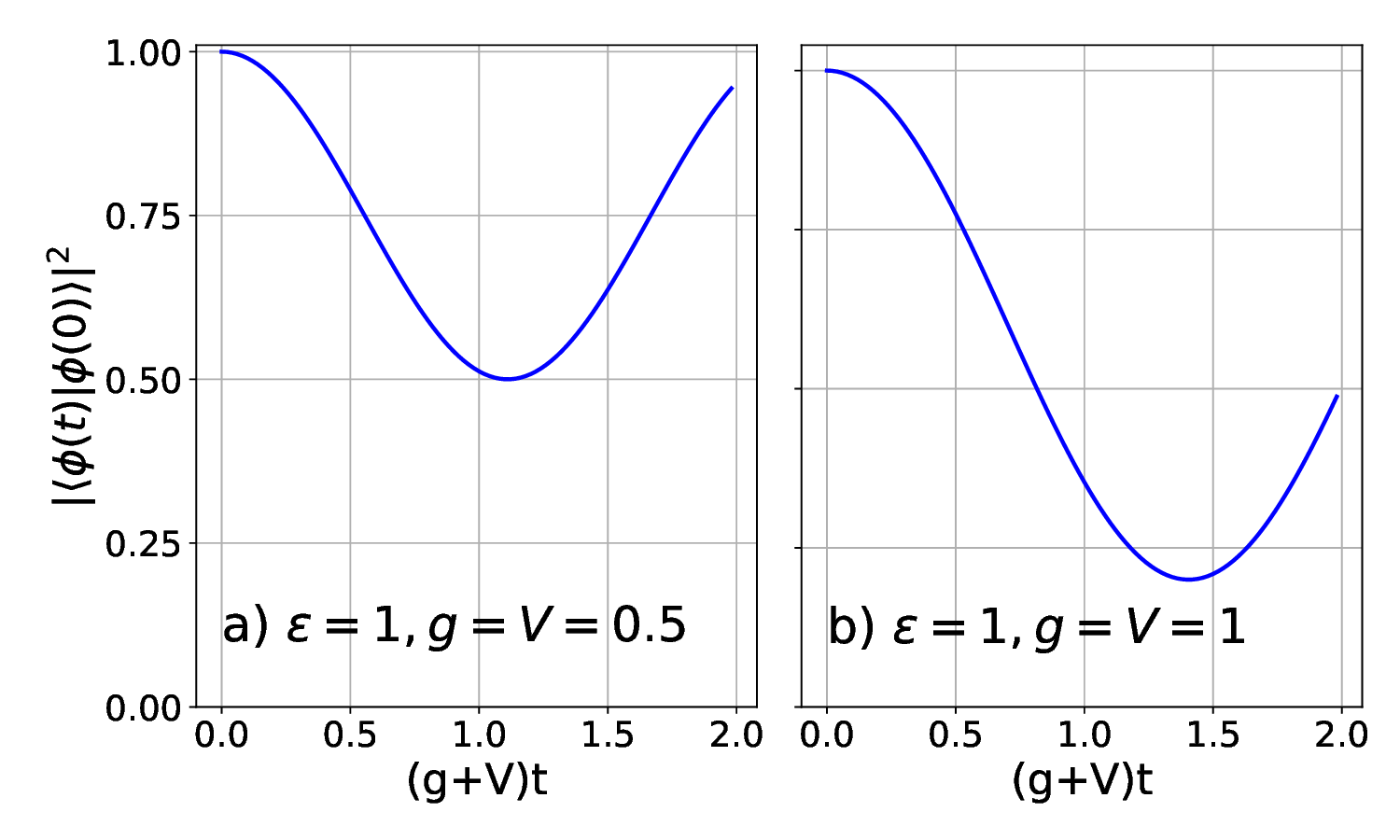}
  \caption{Survival probability $|\langle \phi (t)| \phi(0) \rangle|^2$ as a function of $(g+V) t$ for $\epsilon=1$, $g=V=0.5$ in panel a) and  $\epsilon=1$, $g=V=1$ in panel b). The considered initial state is $|\downarrow_1\otimes\downarrow_2\otimes\uparrow_3\otimes\uparrow_4\rangle$. }
  \label{fig:survival}
\end{figure}
Calculations with larger systems \cite{Saiz21} points into the same conclusions raised in this work, however digital quantum simulations for $j>>1$ are beyond current technology.    

\medskip

\begin{figure}[hbt]
  \centering
  \includegraphics[width=1\linewidth]{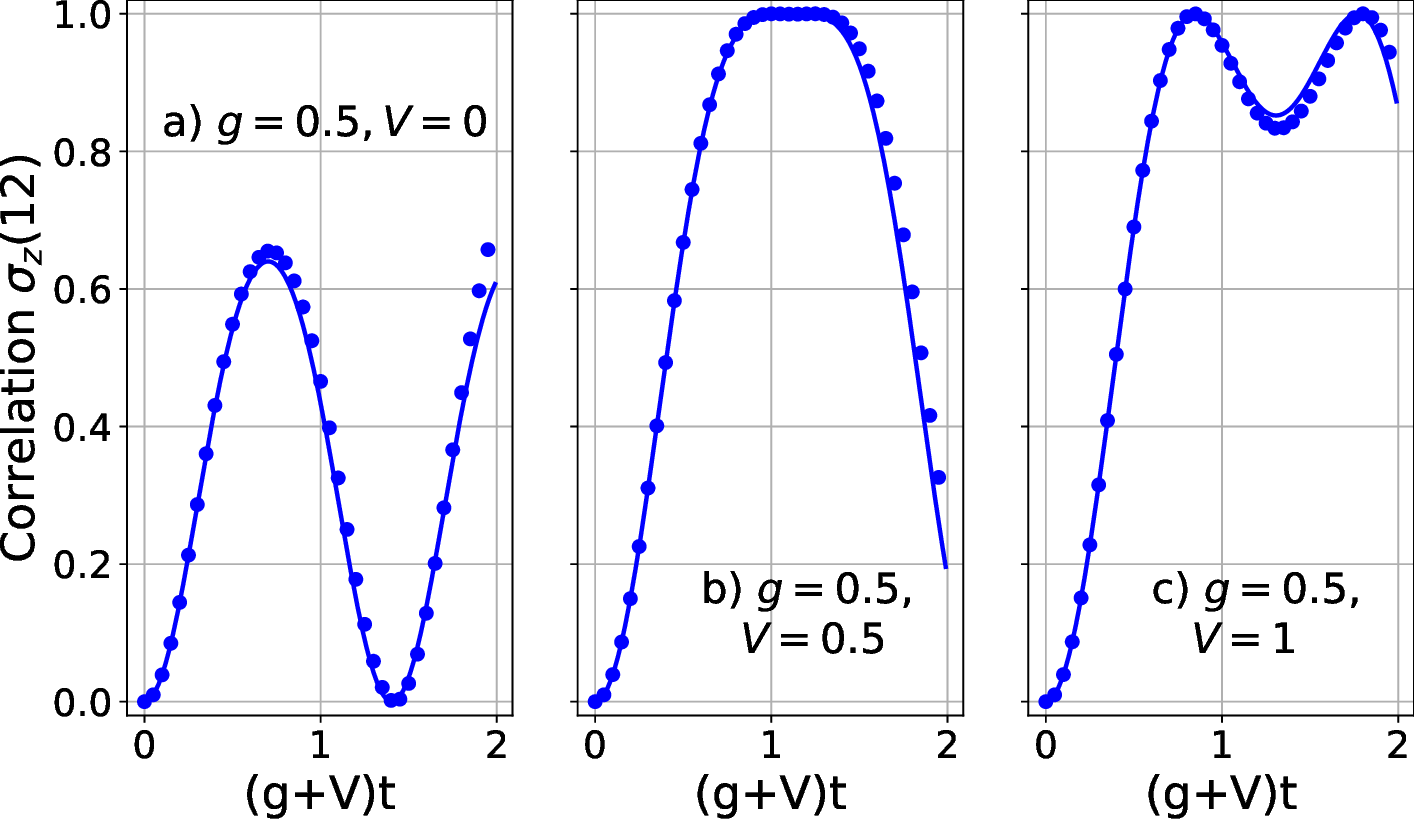}
  \caption{Correlation function $\sigma_z(1,2)\equiv \langle \sigma_1^z \sigma_2^z\rangle-\langle \sigma_1^z \rangle\langle\sigma_2^z\rangle$ for an initial state $|\downarrow_1\otimes\downarrow_2\otimes\uparrow_3\otimes\uparrow_4\rangle$ and Hamiltonian parameters $\epsilon=1$, $g=0.5$, and $V=0$ in a), $\epsilon=1$, $g=0.5$, and $V=0.5$ in b), and $\epsilon=1$, $g=0.5$, and $V=1$ in c). Lines correspond to exact calculations while dots refer to a Trotter expansion with $n_T=5$. 
  }
    \label{fig:correlation}
  \end{figure}

Also, we can observe in Fig.~\ref{fig:fidelity} that the digital error remains negligible for a sizeable time evolution with a nontrivial dynamics and a sufficiently large $n_T$. With respect to the total gate error in a plausible implementation with trapped ions, one can estimate its magnitude via adding the single and two-qubit gate errors times the corresponding number of gates. In our specific 4-qubit proposal, there are 52 single-qubit gates and 50 two-qubits gates. If one assumes experimentally achieved values of 0.0001 for the single-qubit gate error~\cite{DavidLucasSingleQubit} and 0.001 for the two-qubit one~\cite{DavidLucasTwoQubit}, an estimate for the total gate error $E_G$  will be, assuming $n_T=5$, $E_G\simeq 5\times(52\times 0.0001+50\times 0.001)\simeq 0.28$. 
 Thus, with a conservative gate counting, we estimate that the achieved fidelity may be above 70\%. 
 Moreover, the number of gates is such that one may perform the experiment well before the decoherence time, in less than ten milliseconds. Therefore, our proposal may be carried out in trapped-ion setups with current technology, for a proof-of-principle model with $j=1$, i.e., $N=4$. We point out that in this heuristic analysis we have assumed a well-controlled experiment with uncorrelated errors. In addition, the protocol is efficiently scalable to many fermionic modes, namely, $N\gg 1$, once the single and two-qubit gate fidelities, as well as coherence times, are improved. This is due to the fact that the number of terms in our digital decomposition scales polynomially in $N$, as opposed to classical supercomputers, for which the scaling would be exponential in the general case.
 Moreover, the nuclear physics models we consider will always have a polynomial number of terms when expressed as sum of many-body tensor products of Pauli matrices. This is due to the fact that the $H_3$ Hamiltonian will contain at most products of four $c$ fermionic operators, and this implies that the number of $\sigma^+$ and $\sigma^-$ operators will be at most of four per term, independently of the number $N$ of modes.
 Regarding the connection to usual observables in nuclear physics, in a quantum simulation experiment such as this, one may compute the quantum state via quantum tomography~\cite{NielsenChuang}, for systems up to 8 qubits, and the Hamiltonian spectra via quantum phase estimation algorithm~\cite{LloydQPE}, which is polynomial (i.e., efficient) in the size of the system.
We point out that quantum tomography would only be useful for a quantum experiment with few qubits, as the one explicitly described here. For scaling up the experiment to many qubits, we propose to employ instead two-point correlation functions as shown in Fig. \ref{fig:correlation}, which can be measured directly in trapped ions via resonance fluorescence.
%
In nuclear physics, sometimes observables evaluated in different times are desirable. In this sense, we point out that it is possible to carry out this kind of measurement in a digital quantum simulator, as was proposed, e.g., in Ref. \cite{Somm02, Garc17}.
\medskip

\begin{figure}[hbt]
  \centering
  \includegraphics[width=.7\linewidth]{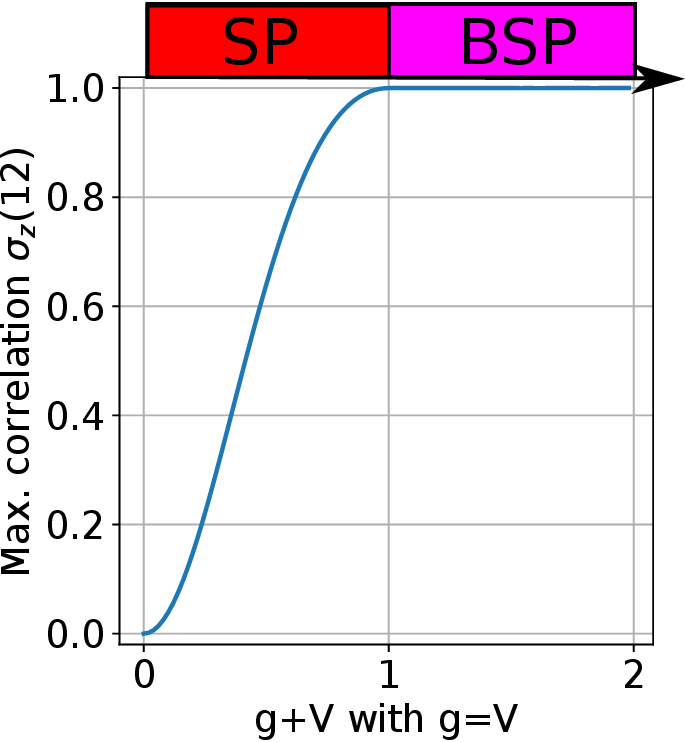}
  \caption{Maximum value of the correlation function $\sigma_z(1,2)$ for an initial state $|\downarrow_1\otimes\downarrow_2\otimes\uparrow_3\otimes\uparrow_4\rangle$ as a function of the Hamiltonian parameters $g=V$ (with $\epsilon=1$). In the upper part the phase diagram is sketched: SP for the symmetric phase and BSP for the broken-symmetry phase. The critical point corresponds to $g=V=0.5$.}
    \label{fig:max_correlation}
  \end{figure}

In Fig.\ \ref{fig:survival} we plot the survival probability $|\langle \phi (t)| \phi(0) \rangle|^2$ as a function of $(g+V) t$ to show that the dynamics of the system is not trivial and significantly changes in the time interval considered in Fig.~\ref{fig:fidelity}. 
Finally, in Fig.~\ref{fig:correlation} we depict the correlation function $\sigma_z(1,2)\equiv \langle \sigma_1^z \sigma_2^z\rangle-\langle \sigma_1^z \rangle\langle\sigma_2^z\rangle$, showing that the time dynamics alone can be used as a probe to explore the different quantum phases of the system via this correlation function. As mentioned above, the critical point in the Agassi model for $j=1$ is given by $g+V=1$. Fig.~\ref{fig:correlation} shows three calculations for this correlation function: a) $g+V<1$ (symmetric phase), b) $g+V=1$ (phase transition point), and c) $g+V>1$ (non-symmetric phase). One can clearly see that at one side of the phase transition the correlation amplitude maximum is smaller than one (symmetric phase, Fig.~\ref{fig:correlation}a with $g+V<1$), it is already one at the transition point (Fig.~\ref{fig:correlation}b, with $g+V= 1$) continue being 1 at the other side (broken phase, Fig.~\ref{fig:correlation}c, with $g+V>1$ ), and extra oscillations appear which amplitudes depend on the control parameter value (also visible in  Fig.~\ref{fig:correlation}c).
A quite similar behaviour is obtained for other initial states and correlation functions, such as $\sigma_z(1,3)$ or $\sigma_z(3,4)$, obtaining also a maximum amplitude for $g+V=1$.
%
This is more clearly shown in  Fig.~\ref{fig:max_correlation}, in which the amplitude of the oscillation is plotted as a function of the control parameter $g=V$, where the critical point is at $g=V=0.5$. The figure shows that this amplitude reaches the value 1 at the critical point and keeps this value in the broken-symmetry phase ($g+V \ge 1$). In the upper part of the figure the phase diagram of the model is sketched, separating the symmetric phase (SP) and the broken-symmetry phase (BSP). This is an evidence that one does not need to compute the ground state to distinguish the different quantum phases in the system. The most direct time dynamics for a typical initial state allows one to obtain signatures of these quantum phases in the amplitude of the time dynamics of the correlation function. 
This type of procedure resembles a dynamical quantum phase transition (see Ref. \cite{Heyl18}), which is a type of quantum phase transition in the time domain. In Ref. \cite{Pueb20}, the author studied the Rabi model through a quench, noticing that its evolution provides hints on the phase of the ground state of the system. 
The analysis of the time evolution of the correlation function $\sigma_z(1,2)$, among other possible functions, provides information on the whole Hilbert space of the system, including its ground state. We plan to extend the present formalism to larger systems and to other observables, such as the Loschmidt echo, to explore how robust are the obtained results.
%
Even very small systems, such as the one we are considering with $j=1$, can present some precursors of quantum phase transitions as it was  explained in Ref. \cite{Iach04}. Although the quantum phase transition is strictly defined in the thermodynamic limit, already for really small sizes its effect is noticeable in several observables that can act as order parameters.

\section{Conclusions}
We have proposed and analyzed the quantum simulation of the Agassi model.
Our numerical simulations and analytical estimations show that this protocol is feasible with current technology, for instance, using trapped ions. The proposal has been exemplified with four sites to be implemented with four trapped ions, while it is scalable to many sites with polynomial resources. We also give evidence that the time dynamics of a quantum correlation function for typical initial states can serve as a probe to explore the different quantum phases of the model, with no need of computing specifically the ground state. Indeed, the different phases of the system can be matched to the time dynamics of the amplitudes of the correlation function.
With recent advances in trapped-ion quantum platforms approaching a few tens of ions in a quantum processor \cite{IonQ,Pogorelov}, we are already going through the crossover for outperforming the fastest classical supercomputers for useful scientific problems. Our approach is a step in this direction, for the efficient quantum simulation of the Agassi model and related nuclear physics systems with
digital quantum platforms. An appeal of trapped-ion quantum platforms is the all-to-all connectivity that enables one to implement the N-body tensor products of Pauli matrices with just two M{\o}lmer-S{\o}rensen gates. However, to be able to carry out a full-fledged quantum simulation of the Agassi model with trapped ions, two-qubit gate fidelities will still need to improve.
%
Even though the scaling of our quantum algorithm is polynomial in the number of quantum particles, beyond a few hundred spins one will need to employ a full fledged error corrected quantum computer, with the consequent overhead in resources that will be needed.

\section*{Acknowledgements}
This work was partially supported by the Consejer\'{\i}a de Econom\'{\i}a, Conocimiento, Empresas y Universidad de la Junta de Andaluc\'{\i}a (Spain) under Groups FQM-160, FQM-177, and FQM-370, and under projects P20-00617, P20-00764,  P20-01247, UHU-1262561, and US-1380840; by grants PGC2018-095113-B-I00, PID2019-104002GB-C21, PID2019-104002GB-C22, and PID2020-114687GB-I00 funded by MCIN/AEI/10.13039/50110001103 and ``ERDF A way of making Europe'' and by ERDF, ref.\ SOMM17/6105/UGR. Resources supporting this work were provided by the CEAFMC and Universidad de Huelva High Performance Computer (HPC@UHU) funded by ERDF/MI\-NE\-CO project UNHU-15CE-2848.

\end{document}